\documentclass[%
preprint,
nofootinbib,
 amsmath,amssymb,
 aps,
]{revtex4-1}

\usepackage{graphicx}
\usepackage{dcolumn}
\usepackage{bm}
\usepackage{graphics}
\usepackage{bm}
\usepackage{amsmath}
\usepackage{epstopdf}
\usepackage{amsmath}
\usepackage{amsfonts}
\usepackage{color}
\usepackage[]{algorithm2e}
 \usepackage{tikz}
 \usepackage{mathtools}
 \usepackage[percent]{overpic}
 \usepackage{float}
 \usepackage{subfig}
 \usepackage{comment}
\let\MYoriglatexcaption\caption
\renewcommand{\caption}[2][\relax]{\MYoriglatexcaption[#2]{#2}}


\renewcommand{\eqref}[1]{(\ref{#1})}
\newcommand{\argmin}{\arg\min}

\begin{document}

\preprint{APS/123-QED}

\title{Non-Markovian SIR epidemic spreading model} 

\author{Lasko Basnarkov$^{1,2}$}
\email{lasko.basnarkov@finki.ukim.mk}
\author{Igor Tomovski$^{2}$}
\author{Trifce Sandev$^{2,3,4}$}
\author{Ljupco Kocarev$^{1,2}$}

\affiliation{
$^{1}$SS. Cyril and Methodius University, Faculty of Computer Science and Engineering,  P.O. Box 393, 1000 Skopje, Macedonia}%
\affiliation{
$^{2}$Macedonian Academy of Sciences and Arts, P.O. Box 428, 1000 Skopje, Macedonia}
\affiliation{
$^{3}$Institute of Physics \& Astronomy Karl-Liebknecht-Str. 24/25,
University of Potsdam, D-14476 Potsdam-Golm, Germany}%
\affiliation{
$^{4}$Institute of Physics, Faculty of Natural Sciences and Mathematics,
Ss Cyril and Methodius University, Arhimedova 3, 1000 Skopje, Macedonia}%
\date{\today}

\begin{abstract}
We introduce non-Markovian SIR epidemic spreading model inspired by the characteristics of the COVID-19, by considering discrete- and continuous-time versions. The incubation period, delayed infectiousness and the distribution of the recovery period are modeled with general functions. By taking corresponding choice of these functions, it is shown that the model reduces to the classical Markovian case. The epidemic threshold is analytically determined for arbitrary functions of infectivity and recovery and verified numerically. The relevance of the model is shown by modeling the first wave of the epidemic in Italy, in the spring, 2020.

\end{abstract}

\maketitle

\section{Introduction}

The ongoing pandemics of COVID-19, has claimed millions of human lives, caused stagnation of the global economy and excessive load on the healthcare systems throughout the world and changed the normal life. Mathematical models of epidemic spreading are important tools for predicting the effects that the pandemics can have on each segment of the society. They provide support for policy-makers to make adequate decisions in order to partially mitigate the consequences by planning various social distancing measures, preparation of healthcare facilities and appropriate adaptation of the economy. 

The spectrum of mathematical models applied for the COVID-19 pandemic ranges from the simplest SIR to rather complex SIDARTHE \cite{roda2020difficult, zhao2020modeling,  calafiore2020time, giordano2020modelling, gatto2020spread}, which are used for assessment of different aspects of the epidemics. One of the major features of these models is their Markovian nature, which considers transitions from one state to another to be independent on the past. As an example, when Markovian property is assumed to hold, an individual that has just become infected can proceed to recovered state with the same probability as another one which has been infected for longer period. This Markovian assumption, encapsulated in constant transition probabilities, or rates, makes the models easier to study analytically. The outcomes of these studies with Markovian approach offer some, and in certain  instances  satisfactory, assessment of the spreading dynamics. However, growing body of evidence, particularly for the COVID-19, suggests existence of incubation  period and certain infectivity patterns, with possibility for spreading the pathogen before onset of the symptoms, to which correspond functions that are rather distinct from the exponential distribution which the Markovian models rely on \cite{Qin2020, qin2020estimation}. Although adding one or more compartments for the Exposed, Asymptomatic, Presymptomatic, or Quarantined persons or considering various kinds of delay \cite{liu2020covid, dell2020solvable, rong2020effect} 
address such observations to certain extent, they cannot systematically incorporate the observed distributions of the incubation period and the healing process.

The non-Markovian setting is inherent in the pioneering works in the mathematical epidemiology by Ross \cite{ross1916application, ross1917application, ross1917application2}, Kermack and McKendrick \cite{kermack1927contribution}, and in the related field of population dynamics by Br\"{o}ck \cite{bockh1877statistisches} and Lotka \cite{lotka1919contribution}. However, the more special and mathematically more tractable, Markovian approach has largely dominated in subsequent studies. In the recent time the non-Markovian framework has started to gain more attention in various settings. In one attempt \cite{boguna1} is proposed Gillespie algorithm as an adequate tool for numerical analysis of non-Markovian spreading models. The effects of the form of distribution of infection and curing (recovery) times on SIS epidemic model occurring on complex networks in continuous time has been analyzed in several studies \cite{starnini,delft_nm1, delft_nm2, delft_nm3, Feng2019, krylova2013effects}. With the introduction of SI*V* model \cite{Nowzari} it was  suggested that non-Markovian spreading models have capacity to be extended to cover a wide variety of spreading sub-models and variants. Nontrivial distribution of infectious period in an integro-differential SIR model was considered in \cite{riano2020epidemic}. In a recent study, non-Markovian SIS model on complex networks, with arbitrary function for infectivity and recovery was proposed \cite{tomovski2021epidemic}, in which control theory was successfully applied for determination of epidemic threshold. Another, novel key contributions in the theory of non-Markovian epidemic spreading models can be considered \cite{pang2020functional, forien2021estimating}. In those works, with extensive theoretical work on models with integro-differential equations were obtained analytical results about the equilibria and the basic reproduction numbers.
Our study adds determination of the epidemic threshold on base on the stability analysis for general distributions of infectivity and healing in a SIR model. By similar approach as in \cite{tomovski2021epidemic} we show how these functions determine the epidemic threshold. The relevance of the model, besides by numerical simulations, is verified by fitting to the observations of the first wave of the epidemic in Italy, in the spring, 2020.

The paper is organized as follows. After providing initial setting of the model in Section \ref{sec:preliminaries}, we introduce the discrete-time and continuous-time models in Sections \ref{sec:discrete} and \ref{sec:cont}, respectively, where we also derive the epidemic threshold relationships. The reduction to Markovian case of the model is presented in Section \ref{sec:Markov}, while numerical simulations and discussions are given in Section \ref{sec:numerics}. The paper concludes with Section \ref{sec:conclusions}.

\section{Preliminaries} \label{sec:preliminaries}

We consider SIR model that has three compartments: Susceptible - S, Infected - I and Recovered - R, with the usual transition $S \to I \to R$. Let the corresponding variables $S$, $I$ and $R$ denote the fractions of the population that are in the given state, and under assumption without births and deaths, one has the normalization condition $S(t)+I(t)+R(t) = 1$ at each moment $t$. To capture the nontrivial dependence of the healing period and the different contagiousness of the infected individual in different stages of the disease we introduce two functions. The infectivity function $\beta(\tau)$ captures the rate, or probability at which individuals that became infected before time $\tau$ are spreading the disease to the susceptible ones. Thus, by simply taking $\beta(\tau) = 0$ for $\tau < T_0$, one is able to introduce incubation period with length $T_0$. Another important function is the healing function $\gamma(\tau)$ that denotes the probability with which individual can heal at moment $\tau$ after contracting the disease. To account for asymptomatic transmitters and existence of certain time window when presence of pathogen can be confirmed, one can introduce a reporting function $\rho(\tau)$. It is associated to the probability that the presence of the pathogen can be confirmed at moment $\tau$ after contraction with it. The asymptomatic cases are conveniently handled by normalizing the reporting function to value smaller than unity. We pursue by considering discrete- and continuous-time models separately, and provide more details about these functions.

\section{Discrete-time version}
\label{sec:discrete}
In this section we consider evolution in discrete time $t$ and denote the fraction of individuals that have become infected within the continuous-time interval $[t-1, t]$ with $I_d(t)$, where for simplicity the unit interval is taken to be 1. This can be relevant for situations like those when cases are considered on daily basis. In such scenario, we have discrete-time healing function $\gamma(\tau)$ and infectivity one $\beta(\tau)$, on which we put the constraint $\beta(0) = 0$. The probability that the individual will heal within $\tau$ time units is $\Gamma(\tau) = \sum_{\nu=0}^{\tau} \gamma(\tau)$. We further assume finite duration $T$ of the disease, what implies $\Gamma(T) = 1$ and for practical reasons introduce its complement $\overline{\Gamma}(\tau) = 1 - \Gamma(\tau)$, to denote the probability that individual has not healed yet for $\tau$ time units. The function $\gamma(\tau)$ also has a meaning of fraction of individuals that have contracted the disease within the same unit time interval, to become healed later within another unit interval $[\tau-1, \tau]$. Similar reasoning holds for the cumulative functions $\Gamma(\tau)$ and $\overline{\Gamma}(\tau)$. On base on the classical SIR model, the proposed model of evolution of the compartments is given with the system
\begin{eqnarray}
    S(t + 1) &=& S(t) \left[1 - \sum_{\tau=0}^{T-1} \beta(\tau) \overline{\Gamma}(\tau) I_d(t-\tau)\right] \nonumber \\
    I_d(t + 1) &=& S(t) \sum_{\tau=0}^{T-1} \beta(\tau)\overline{\Gamma}(\tau) I_d(t-\tau) \nonumber \\
    R(t + 1) &=&  R(t) +\sum_{\tau=0}^{T-1} \gamma(\tau) I_d(t + 1 -\tau). \label{eq:discrete_model}
\end{eqnarray}

One can note that the infected individuals that have contracted the pathogen up to $T$ periods before the current moment $t$, and which are not healed yet, can contribute to spreading of the disease, with appropriate intensity captured in the function $\beta(\tau)$. We note that in order to determine the infected fraction at given moment, one should sum those infected in the past, but did not heal up to the given moment
\begin{equation}
    I(t) = \sum_{\tau=0}^{T-1} I_d(t-\tau) \overline{\Gamma}(\tau).
    \label{eq:discrete_total_inf}
\end{equation}
To make the problem completely defined one has to specify the initial conditions for $I_d(t)$. We assume that they are given for $\tau = T-1, T-2, \dots, 0$. In general this model cannot be solved analytically and should be studied by application of numerical simulations.

To get insight of the conditions when epidemic can emerge, one can determine the stability of the disease free state $S^*=1, I^*=I_d^* = R^*=0 $, that is an equilibrium point of the system. Its local stability is established by linearizing the dynamical equations (\ref{eq:discrete_model}) in its neighborhood. By making the linearization in vicinity of $S^*=1, I^*=R^*=0$, one can observe the dynamical evolution of the perturbations $\delta S = S - S^*, \delta I_d =I_d - I_d^*, \delta R = R- R^*$. Under linearization, the perturbations are related with
\begin{eqnarray}
    \delta S(t+1) &=& \delta S(t) -\sum_{\tau=0}^{T-1} \beta(\tau) \overline{\Gamma}(\tau) \delta I_d(t-\tau), \nonumber \\
    \delta I_d(t+1) &=& \sum_{\tau=0}^{T-1} \beta(\tau)\overline{\Gamma}(\tau) \delta I_d(t-\tau), \nonumber\\ 
    \delta R(t + 1) &=& \delta R(t) + \sum_{\tau=0}^{T-1} \gamma(\tau) \delta I_d(t + 1-\tau). \label{eq:discrete_model_pert}
\end{eqnarray}

Let us focus on the infected fraction and make $Z$-transform on the second equation in (\ref{eq:discrete_model_pert}). To do so, multiply first both sides of that equation by $z^{-t}$ and sum to obtain
\begin{equation}
    \sum_{t=0}^{\infty} \delta I_d(t+1) z^{-t} = \sum_{t=0}^{\infty} \sum_{\tau=0}^{T-1} \beta(\tau)\overline{\Gamma}(\tau) \delta I_d(t-\tau) z^{-t}. \label{eq:Z_transform_disc}
\end{equation}
By using the $Z$-transform of the fraction of the population that become infected at unit interval $I_d(t)$, given as $\mathcal{I}(z) = \sum_{t=0}^{\infty} I_d(t) z^{-t}$, the left hand side of (\ref{eq:Z_transform_disc}) will become
\begin{eqnarray}
    \sum_{t=0}^{\infty} \delta I_d(t+1) z^{-t} &=& z\sum_{t=0}^{\infty} \delta I_d(t+1) z^{-(t+1)}\nonumber\\ &=& z\left[\mathcal{I}(z)- \delta I_d(0)\right]. \label{eq:delta_ID_t_left}
\end{eqnarray}
Accordingly, the right-hand side of (\ref{eq:Z_transform_disc}) can be rearranged as 
\begin{eqnarray}
    &&\sum_{t=0}^{\infty} \sum_{\tau=0}^{T-1} \beta(\tau)\overline{\Gamma}(\tau) \delta I_d(t-\tau) z^{-t} =\nonumber\\ =&&\sum_{\tau=0}^{T-1} \beta(\tau)\overline{\Gamma}(\tau) z^{-\tau} \sum_{t=0}^{\infty} \delta I_d(t-\tau) z^{-(t-\tau)}.
\end{eqnarray}
By using substitution $\nu = t-\tau$, the last sum for $\tau \leq -1$ can be expressed as
\begin{equation}
    \sum_{\nu=-\tau}^{\infty} \delta I_d(\nu) z^{-\nu} =\sum_{\nu=-\tau}^{-1} \delta I_d(\nu) z^{-\nu} + \mathcal{I}(z) =
    \mathcal{I}_{0}(\tau, z) + \mathcal{I}(z), \label{eq:delta_ID_t_right}
\end{equation}
where we have introduced a function $\mathcal{I}_0(\tau, z)$ that corresponds to the initial conditions. Now, combining the relationships (\ref{eq:delta_ID_t_left}) -- (\ref{eq:delta_ID_t_right}) one has
\begin{equation}
    z\left[\mathcal{I}(z)- \delta I_d(0)\right]= \sum_{\tau=0}^{T-1} \beta(\tau)\overline{\Gamma}(\tau) \left[\mathcal{I}_0(\tau, z) + \mathcal{I}(z)\right] z^{-\tau} .
\end{equation}
To shorten the notation, one can introduce the following two complex functions
\begin{eqnarray}
    \mathcal{E}(z) &=& \sum_{\tau=0}^{T-1} \beta(\tau)\overline{\Gamma}(\tau) z^{-\tau}, \nonumber\\
    \mathcal{E}_0(z) &=& \sum_{\tau=0}^{T-1} \beta(\tau)\overline{\Gamma}(\tau) \mathcal{I}_0(\tau, z) z^{-\tau}.
\end{eqnarray}
The first one is simply the $Z$-transform $\mathcal{E}(z)$ of what might be called epidemic function $E(\tau) = \beta(\tau)\overline{\Gamma}(\tau)$, that is a combination of the infecting and healing functions because $\sum_{\tau=0}^{T-1} \beta(\tau)\overline{\Gamma}(\tau) z^{-\tau}=\sum_{\tau=0}^{\infty} \beta(\tau)\overline{\Gamma}(\tau) z^{-\tau}$. The second complex function $\mathcal{E}_0(z)$ is related to the initial conditions. Now, one has the following relationship
\begin{equation}
    z\left[\mathcal{I}(z)- \delta I_d(0)\right] = \mathcal{I}(z)\mathcal{E}(z) + \mathcal{E}_0(z),
\end{equation}
from where 
\begin{equation}
    \mathcal{I}(z) = \frac{z\delta I_d(0)+\mathcal{E}_0(z)}{z-\mathcal{E}(z)}.\label{eq:infected_Z_transform}
\end{equation}
From a result in theory of discrete linear time-invariant systems, a sequence (the impulse response of such system) is decaying if the poles of its $Z$-transform are within the unit circle \cite{oppenheim2013signals}. Thus, when the poles of the function $\mathcal{I}(z)$ of the complex function (\ref{eq:infected_Z_transform}), or the roots of the polynomial $z-\mathcal{E}(z)$ lie within the unit circle, the perturbation dies out at infinity. So, the epidemic threshold can be obtained by taking $z=1$ in the denominator in (\ref{eq:infected_Z_transform}), that results in
\begin{equation}
    \sum_{\tau=0}^{T-1} \beta(\tau)\overline{\Gamma}(\tau) = 1, \label{eq:threshold_discrete}
\end{equation}
which obviously depends on the functional forms of the infectivity and healing functions.

We should finally note that any initial infection would not shift back the population to the disease-free state $S=1, I=R=0$, but to some endemic $S_e^*, I_e^* = 0, R_e^* = 1-S^*$. However, if the conditions are not favoring epidemic both equilibria will be rather close $S_e^* \approx 1$.

\section{Continuous-time version}\label{sec:cont}

We will pursue similarly to the discrete-time approach, where the fractions of individuals within given compartment and the functions modeling the infectivity, healing and reporting are defined for continuous time $t$ and we use the same notation. Thus, $S(t)$ is the fraction of susceptible individuals at given moment $t$ and $R(t)$ corresponds to the recovered and again assume finite healing period $T$. The fraction of infected individuals is conveniently modeled with the rate of infection, or the fraction of newly infected individuals $I_d(t)$ within the infinitesimal interval $(t-dt, t)$. The total fraction of infected persons is given with the integral
\begin{equation}
    I(t) = \int_{0}^{T} I_d(t-\tau) \overline{\Gamma}(\tau) d\tau, \label{eq:total_I}
\end{equation}
which accounts for those that had become infected in the past and have not healed yet. Now, the dynamical evolution of the respective fractions is given with
\begin{eqnarray}
    \dot{S} &=& - S(t)\int_{0}^{T} \beta(\tau) \overline{\Gamma}(\tau) I_d(t-\tau)d\tau \nonumber \\
    I_d(t) &=& S(t) \int_{0}^{T} \beta(\tau)\overline{\Gamma}(\tau)  I_d(t-\tau) d\tau \nonumber \\
    \dot{R} &=& \int_{0}^{T} \gamma(\tau) I_d(t-\tau) d\tau. \label{eq:continuous_model}
\end{eqnarray}
One should note that in their original approach, the general version of the model by Kermack and McKendrick assumes dependence of the infectivity on the age of infection just as the last relationships (\ref{eq:continuous_model}) suggests \cite{kermack1927contribution, brauer2017mathematical}.
In order to determine whether the initial perturbation will grow to epidemics, one could focus on the second equation in the vicinity of the disease-free state $S^*=1, R^*=I^*=0$. Then, the perturbation of newly infected individuals will evolve as 
\begin{equation}
   \delta I_d(t) = \int_{0}^{T} \beta(\tau)\overline{\Gamma}(\tau) \delta I_d(t-\tau)d\tau, \label{eq:pert_infect_rate}
\end{equation}
where it is assumed that in vicinity of the disease-free state $S(t)\approx 1$. Now, make Laplace transform of the perturbation of the rate of infection, $\mathcal{I}(s) = \int_0^{\infty}\delta I_d(t)e^{-st}dt$ and use it in the last equation (\ref{eq:pert_infect_rate}). To do that, we will follow the same approach as in the discrete-time version. Multiply both sides with $e^{-st}$ and integrate. The left hand side will result in the Laplace transform of $\delta I_d(t)$, while the right hand one will be 
\begin{eqnarray}
    A &= &\int_0^{\infty}\int_{0}^{T} \beta(\tau)\overline{\Gamma}(\tau) \delta I_d(t-\tau)e^{-st}d\tau dt \nonumber\\ &=& \int_{0}^{T} \beta(\tau)\overline{\Gamma}(\tau) e^{-s\tau} \int_0^{\infty} \delta I_d(t-\tau) e^{-s(t-\tau)} dt \nonumber \\
    &=&\int_{0}^{T} \beta(\tau)\overline{\Gamma}(\tau) e^{-s\tau} \int_{-\tau}^{\infty}\delta I_d(\nu) e^{-s\nu} d\nu
\end{eqnarray}
The last integral can be expressed with
\begin{eqnarray}
\int_{-\tau}^{\infty}I_d(\nu) e^{-s\nu} d\nu &=& \int_{-\tau}^{0}I_d(\nu) e^{-s\nu} d\nu + \mathcal{I}(s)\nonumber\\
&=& \mathcal{I}_0(\tau,s) + \mathcal{I}(s).
\end{eqnarray}
Now, one has
\begin{equation}
    A=\int_{0}^{T} \beta(\tau)\overline{\Gamma}(\tau) e^{-s\tau} \left[\mathcal{I}_0(\tau,s) + \mathcal{I}(s)   \right] d\tau.
\end{equation}
Similarly to the discrete-time case we can introduce the Laplace transform of the epidemic function $E(\tau)=\beta(\tau)\overline{\Gamma}(\tau)$ and its initial conditions contribution
\begin{eqnarray}
    \mathcal{E}(s) &=& \int_{0}^{T} \beta(\tau)\overline{\Gamma}(\tau) e^{-s\tau} d\tau, \\ \nonumber
    \mathcal{E}_0(s) &=& \int_{0}^{T} \beta(\tau)\overline{\Gamma}(\tau) \mathcal{I}_0(\tau,s) e^{-s\tau} d\tau.
\end{eqnarray}
Finally, one obtains
\begin{equation}
    \mathcal{I}(s) =  \mathcal{I}(s) \mathcal{E}(s) + \mathcal{E}_0(s),
\end{equation}
from where the Laplace transform of the perturbation of the infection rate is
\begin{equation}
    \mathcal{I}(s) = \frac{\mathcal{E}_0(s)}{1-\mathcal{E}(s)}. \label{eq:Laplace_transf_pert}
\end{equation}
From the results of control theory, a continuous-time linear time-invariant system is stable if the poles of its transfer function, or Laplace transform of its impulse response have negative real part \cite{oppenheim2013signals}. Thus, the perturbations $\delta I_d(t)$ will decay if the poles of its Laplace transform $\mathcal{I}(s)$ (\ref{eq:Laplace_transf_pert}), or eigenvalues of the system (\ref{eq:continuous_model}) lie within negative half-plane $Re\{s\} < 0$. Then, the epidemic threshold can be obtained with $s=0$ which leads to
\begin{equation}
    \int_0^T \beta(\tau)\overline{\Gamma}(\tau) d\tau = 1, \label{eq:Threshold_continuous}
\end{equation}
that represents the relationship, which corresponds to the discrete-time case (\ref{eq:threshold_discrete}).


\section{Markovian SIR model}\label{sec:Markov}

In order to obtain the classical Markovian SIR model, from the non-Markovian case (\ref{eq:discrete_model}), one should consider taking $T \rightarrow \infty$, $\beta(\tau)=\beta$, $\gamma(0)=0$, $\gamma(\tau)=\gamma(1-\gamma)^{\tau-1}$ where $\beta$ and $\gamma$ are constants. This further yields $\Gamma(0)=0$, $\Gamma(\tau)=1-(1-\gamma)^{\tau}$ and $\overline{\Gamma}(\tau)=(1-\gamma)^{\tau}=\gamma(\tau+1)/\gamma$. First, one could observe that by using constant infectivity $\beta(\tau)=\beta$ in the first relationship of the model (\ref{eq:discrete_model}) and using (\ref{eq:discrete_total_inf}) one will obtain the classical form for evolution of the susceptible population
\begin{equation}
    S(t + 1) = S(t) \left[1 - \beta I(t)\right]. \label{eq:SIR_S_classical}
\end{equation}
Next, by implementing the condition $\gamma(0) = 0$, and the relationship $\overline{\Gamma}(\tau)=\gamma(\tau+1)/\gamma$ one can drop the first term in the sum in the recovered population in (\ref{eq:discrete_model}), and further obtain
\begin{eqnarray}
    &&\sum_{\tau=0}^{T-2} \gamma(\tau+1) I_d(t-\tau) \nonumber\\ &&= \gamma\sum_{\tau=0}^{T-1} \overline{\Gamma}(\tau) I_d(t-\tau)- \gamma \overline{\Gamma}(T-1) I_d(t-T+1)\nonumber\\ &&= \gamma I(t)- \gamma (1-\gamma)^{T-1}I_d(t-T+1), 
\end{eqnarray}
from where, for $T \rightarrow \infty$, the recovered population evolves as
\begin{equation}
    R(t+1) = R(t) + \gamma I(t). \label{eq:SIR_R_classical}
\end{equation}
Finally, from the conservation relationship $I(t) + S(t) + R(t) = 1$, one can find that the infected fraction is given as
\begin{equation}
    I(t+1) = \beta S(t)I(t)+(1-\gamma) I(t).
    \label{eq:SIR_I_classical}
\end{equation}
The relationships (\ref{eq:SIR_S_classical}), (\ref{eq:SIR_R_classical}) and (\ref{eq:SIR_I_classical}) represent the classical SIR model in discrete time.

\begin{figure}[h]
\includegraphics[width=0.6\columnwidth]{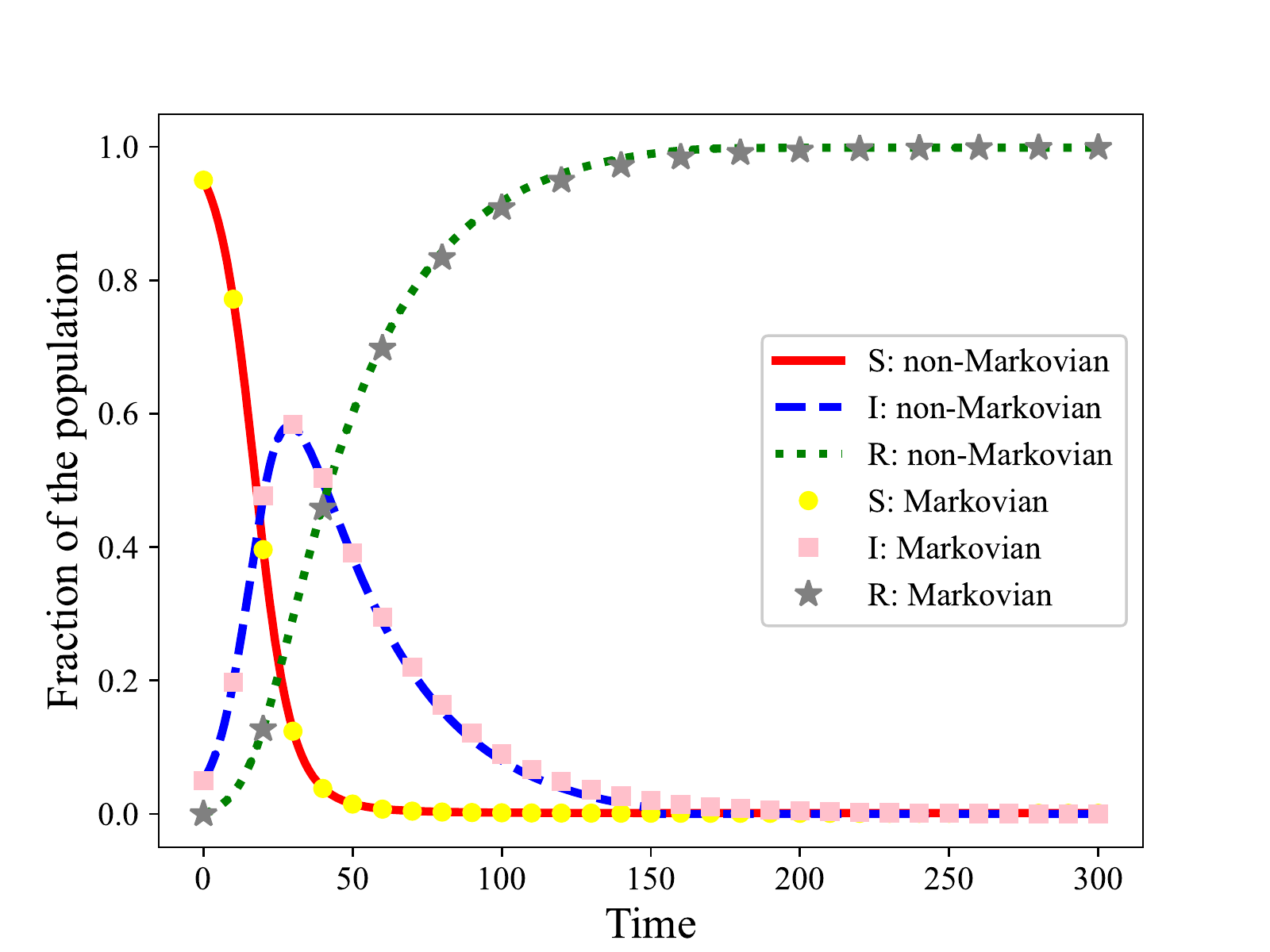}
\centering
\caption{Comparison between the discrete classical SIR model and the classical SIR - equivalent model obtained from the non-Markovian form, for $\beta=0.2$, $\gamma=0.03$. It is used rather large finite duration of the healing $T=150$, as a proxy for $T\to\infty$. }
\label{Mar_vs_nonMar}
\end{figure}

As an example, in the figure \ref{Mar_vs_nonMar} we make a comparison between numerical solutions of the discrete classical SIR model and the classical SIR - equivalent model obtained from the non-Markovian form.

Similarly to the discrete-time version, to verify that the proposed continuous model is generalization of the classical, Markovian SIR model, one should consider two characteristics of the latter: 1. The infection rate is independent on the moment when the disease was contracted $\beta(\tau) = \beta$; and 2. The duration of infectivity is infinite and exponentially distributed which implies that the healing function is $\gamma(\tau) = \lambda e^{-\lambda \tau}$. We note that the respective cumulative distribution is $\Gamma(\tau) = 1 - e^{-\lambda\tau}$, and accordingly $\overline{\Gamma}(\tau) = e^{-\lambda\tau}$. By using the functional form of the healing function, the total infectious population will be
\begin{equation}
    I(t) = \int_0^{\infty} \overline{\Gamma}(\tau)I_d(t-\tau)d\tau = \int_0^{\infty} e^{-\lambda \tau}I_d(t-\tau)d\tau.
    \label{eq:I_total_classical}
\end{equation}
Similarly, by using $\beta(\tau) = \beta$, for the dynamics of the susceptible fraction one has
\begin{equation}
    \dot{S} = -S(t)\beta \int_0^{\infty} e^{-\lambda\tau} I_d(t-\tau) d\tau = -\beta S I,
    \label{eq:S_classical}
\end{equation}
that represents the corresponding relationship in the classical SIR model. Furthermore, by applying the functional form for the healing function, the dynamics of the recovered population will be as follows
\begin{eqnarray}
    \dot{R} = \int_0^{\infty} \lambda e^{-\lambda \tau} I_d(t-\tau) d\tau = \lambda I,
    \label{eq:R_classical}
\end{eqnarray}
that is the respective relationship in the classical SIR model. Finally, by using the conservation principle $S(t)+I(t)+R(t) = 1$, the total infectious fraction will evolve as
\begin{equation}
    \dot{I} = -\dot{S} - \dot{R} = \beta SI - \lambda I,
    \label{eq:I_classical}
\end{equation}
that is the remaining familiar relationship from the classical case. As a final note, we just mention that using respective forms for the infectivity and recovery functions for the Markovian case in the epidemic threshold relationships (\ref{eq:threshold_discrete}) and (\ref{eq:Threshold_continuous}), one will obtain the familiar threshold $\beta_{\mathrm th} = \gamma$. 





\section{Numerical experiments and discussion}\label{sec:numerics}

Our numerical experiments with the proposed model were based on solution of the integro-differential equations for the continuous-time case. We have used the Euler method with step $\Delta t = 0.01$. Although in the model can be used arbitrary functions of infection and recovery, we have chosen to use those that can been found in the literature as appropriate for the COVID-19 pandemic. As suggested in \cite{Qin2020} the infectivity function $\beta(\tau)$ is conveniently represented with Weibull probability density function, with parameters $\alpha=2.04$ and $\lambda=0.103$, which is further truncated to 35 days and normalized. The daily recovering probabilities were modeled with log-normal probability density function $L(\tau; \mu; \sigma) = 1/(\tau\sigma \sqrt {2\pi }) \exp(-\left(\ln \tau-\mu \right)^{2}/(2\sigma ^{2}))$, with parameters $\mu =\ln (\mu _{X}^{2}/(\sqrt {\mu _{X}^{2}+\sigma _{X}^{2}})$, $\sigma ^{2}=\ln \left (1+\sigma _{X}^{2}/\mu _{X}^{2}\right)$ chosen to match a mean value of $\mu _{X}=21$ and standard deviation $\sigma _{X}=6$. The distribution is then normalized to 61 days, and time-shifted for 4 days in order to exclude immediate recovery. This results in the healing function $\gamma(\tau)$ with mean recovery time of $25 \pm 6$ days, in the following fashion
\begin{eqnarray}
\gamma(\tau)= \begin{cases}
    \frac{L(\tau-4; \mu; \sigma)}{\int_{0}^{61}L(\tau; \mu; \sigma)d\tau},& 4\le \tau \le 65,\\
    0,& \text{otherwise}.
\end{cases}
\end{eqnarray}

This construct was based on the results from \cite{sreevalsan2020analysis,  faes2020time}, assuming that: 1. Onset of symptoms (on average) occurs after four days (the time shift); 2. It takes another 7-10 days from onset of symptoms to diagnosis confirmation and hospitalization; 3. Another 10-11 days, on average, are needed from hospitalization to recovery. The period of $T=65$ days is considered in order to include even most extreme cases in which hospitalization exceeded 40 days.

\begin{figure}[h]
\includegraphics[width=0.6\columnwidth]{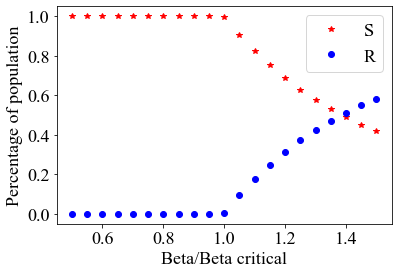}
\centering
\caption{Fractions of susceptible (red stars) and recovered (blue dots) individuals at the end of the epidemic as a function of the scaling of the infectivity function $\beta$ given in terms of its threshold value $\beta_{\mathrm{th}}$.}
\label{fig:Threshold}
\end{figure}

Furthermore, we have chosen to scale the infectivity function with a parameter $\beta$ given in terms of the epidemic threshold $\beta_{\mathrm{th}}$. The threshold value was obtained from the condition (\ref{eq:Threshold_continuous})
\begin{equation}
    \beta_{\mathrm{th}}\int_0^T \beta(\tau)\overline{\Gamma}(\tau) d\tau = 1. \label{eq:beta_critical}
\end{equation}
To verify the value of the epidemic threshold we have varied the infectivity parameter in vicinity of the critical value obtained from (\ref{eq:beta_critical}) and run the continuous-time model for total time equal to 5000. The final values of the susceptible and recovered fraction are plotted as function of the infectivity parameter in the figure \ref{fig:Threshold}. As one can see, once $\beta$ is larger than its critical value, the epidemic emerges.

In order to verify how well the approach can be used to model the COVID-19 pandemic we have chosen to use value of the infectivity parameter $\beta$ that nearly matches the growth patterns of the epidemic in the countries before countermeasures were applied. As was obtained in a detailed study \cite{pellis2020challenges}, the epidemic doubling time in many countries is approximately three days. For that reason, we have opted to use the value $\beta = 4.85\beta_{\mathrm{th}}$ that produces such growth. We have numerically verified that in the initial stage of the epidemic, the newly confirmed daily cases and the total number of infected individuals grow with the same rate, and have the same doubling time. Also, by running the model with $\beta = 4.85\beta_{\mathrm{th}}$ for very long time, we have obtained that at the end less than 1\% of the population will remain susceptible! This result means that, if the doubling time is three days in  case of free spreading of the virus, then prevention of the epidemic would need nearly everyone should be either vaccinated or had healed from the virus. This is particular challenge of the model that should be addressed carefully.

\begin{figure}[h]
\includegraphics[width=0.6\columnwidth]{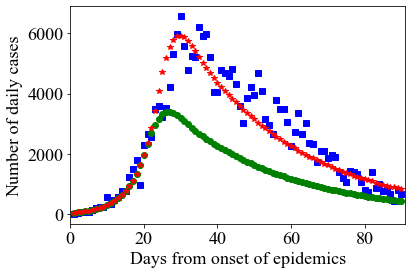}
\includegraphics[width=0.6\columnwidth]{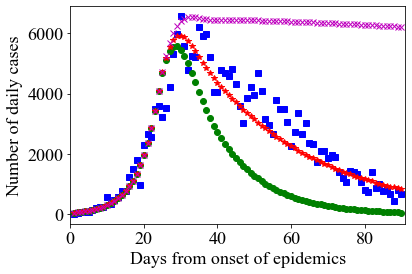}
\centering
\caption{Daily confirmed cases in the first epidemic wave in Italy in spring 2020 (in blue squares), compared to numerical simulations of the model. Top panel: Confirmation function is delayed for two days after onset of infectivity (green circles) and five days (red stars) and $\beta = 0.75\beta_{\mathrm{th}}$. Bottom panel: Confirmation function is delayed for five days, while the infectivity parameter is: $\beta = 0.5\beta_{\mathrm{th}}$ (green circles), $\beta = 0.75\beta_{\mathrm{th}}$ (red stars), and $\beta = \beta_{\mathrm{th}}$ (magenta crosses)}
\label{fig:Italy}
\end{figure}

We have finally attempted to check how well the model can explain the observations. To do so, we have used the COVID-19 data from Our World in Data, for Italy. Our focus was put on the first wave of the pandemic, since in its beginning no preventive measures were used. We have chosen to study the epidemic in Italy, where the wave was the strongest. The window of data under study starts from February 21, 2020, that is the date from which every day were reported new cases. The countrywide lockdown started on March 10, 2020, that corresponds to day 19 in this study. We have used value of infectivity $\beta \approx 3.7\beta_{\mathrm{th}}$ that provided good fit to the observed data for the period from February 21 until March 9. This value was used until the start of the lockdown, when it was set to certain value smaller than the threshold. The initial condition was set to $I_d(0) = 10^{-7}$, that for Italy would mean about 6 persons infected at the starting day of simulation. We have chosen to apply detection of the infected individuals on based on a function that has identical form as the infectivity one, but which is delayed for certain number of days. This corresponds to situation that only those with symptoms are tested, and their appearance is delayed few days after the onset of infectivity. Also, there is certain delay that corresponds to the whole process from onset of symptoms, to visit to hospital to obtaining positive result. We note that the testing function was normalized to 0.8 that corresponds to assuming existence of 20\% asymptomatic cases  \cite{buitrago2020occurrence}. To reach a good fit to the observations we had to take the start of the simulation, that is the day when the initial seed of infection was set, to be approximately 60 days before the day 1, when comparison with the real data starts. Its exact value was obtained by fitting the logarithms of the daily detected cases from the simulation to the respective ones from the data. More precisely, we have looked for a shift $s$, that will result in minimal squared error as follows
\begin{equation}
    \epsilon = \argmin_{s} \left\{\frac{1}{19} \sum_{k=1}^{19} \left[\ln(I_d^{\mathrm{data}}(k)) - \ln(I_d(k+s)) \right]\right\}.
\end{equation}
We report in the top panel of figure \ref{fig:Italy} two simulated scenarios compared to the observations. In the first case we took testing function that is delayed after the infectivity one for two days, that actually becomes nonzero at the possible onset of the symptoms \cite{he2020temporal}, while the other case corresponds to delay of five days. The latter scenario provides much better fit to the observations, particularly in the period after the lockdown has started, and even further in the period after the peak, as one can notice in the figure \ref{fig:Italy}. We have tried with all integer values of the delay from two to ten (not shown) and five days correspond to the best fit. We remind that the lockdown corresponds to day 19 in the plot, while the peaks of the daily reported cases are delayed: at day 27 and day 30, for scenario one, and two, respectively. The peak at the latter case, appears at March 21, the day when largest number of new cases were registered. This fit to the peak and beyond of the model simulation with the observation, makes a good basis for the relevance of the proposed framework. In the bottom panel in figure \ref{fig:Italy} we show how modification of the value of infectivity parameter $\beta$ during the lockdown phase influences the daily cases.

Although providing natural framework for incorporation of observed distributions of the infectiousness of the infected individuals and the typical development of the disease, the proposed model has drawbacks as well. First, before using it, one needs to specify the functions modeling the infectiousness, healing and discovering the infected individuals. Their determination is a serious issue by their own and needs careful study. As more complex one, the tuning of the model would need in general more data than the classical Markovian counterparts. Also, its full specification needs providing initial conditions that represent a high-dimensional vector, or an interval of values. How all these factors shape the outcome of the model, and how much is it robust to perturbations of any kind is unknown. We believe that their understanding could provide the epidemiologists with valuable information for better understanding of the possible outcomes of epidemics with pronounced non-Markovian nature.

\section{Conclusions}\label{sec:conclusions}

The proposed general non-Markovian epidemic spreading model captures the typical patterns of the disease in person infected with SARS-CoV-2: delayed onset of symptoms and infectivity and impossibility of immediate cure of those that will become sick. We have studied both discrete- and continuous-time versions and derived analytically the relationships for determination of the epidemic threshold. The model reduces to the classical SIR model with the corresponding choice of the functions of infection and healing. The theoretical analysis was supported by numerical confirmation of the epidemic threshold values. The good fit of the model to the real data shows its promising potential for application for modeling the spread of other infectious diseases. By introducing other appropriate functions one could possibly generalize this model to versions that include other compartments that correspond to hospitalized, quarantined, or deceased persons. 

Although the epidemic threshold as key quantity was determined, we did not calculated the basic reproduction number $R_0$, that represents another important quantity. Furthermore, the relationship between the scaling of the infectivity function $\beta/\beta_{\mathrm{th}}$ from one side and $R_0$ and the doubling time, from another should be explored as well. With this regard, we think that it is even more important to determine the herd immunity level needed to prevent the epidemic. Finally, analysis of epidemic spreading by nontrivial contact patterns, modeled with complex networks, and by incorporating the proposed approach could provide further insight in the evolution of the epidemics. These issues could provide better understanding of the non-Markovian setting in modeling the epidemic spreading. 

\section{Acknowledgement}
This research was partially supported by the Faculty of Computer Science and Engineering, at the Ss. Cyril and Methodius University in Skopje, Macedonia. The Authors acknowledge support by the German Science Foundation (DFG, Grant number ME 1535/12-1).



%

\end{document}